\documentclass[twocolumn,showkeys,showpacs,preprintnumbers,prd,superscriptaddress,nofootinbib]{revtex4-1}
\usepackage{graphicx}
\usepackage{epsf}
\usepackage{bm}
\usepackage{amsmath}
\usepackage{amsfonts}
\usepackage{amssymb}
\usepackage{epstopdf}
\usepackage{natbib}
\usepackage{hyperref}
\usepackage{color}
\usepackage{verbatim}
\usepackage{multirow}
\usepackage{bm}
\usepackage{hyperref}
\usepackage{listings}
\usepackage{soul}
\begin{document}

\title{Testing beyond-Kerr spacetimes with GWTC-3}

\author{Rafael M. Santos}
\email{rafael.mancini@inpe.br}
\affiliation{Divis\~ao de Astrof\'isica, Instituto Nacional de Pesquisas Espaciais, Avenida dos Astronautas 1758, S\~ao Jos\'e dos Campos, 12227-010, SP, Brazil}

\author{Rafael C. Nunes}
\email{rafadcnunes@gmail.com}
\affiliation{Instituto de F\'{i}sica, Universidade Federal do Rio Grande do Sul, 91501-970 Porto Alegre RS, Brazil}
\affiliation{Divis\~ao de Astrof\'isica, Instituto Nacional de Pesquisas Espaciais, Avenida dos Astronautas 1758, S\~ao Jos\'e dos Campos, 12227-010, SP, Brazil}

\author{Jose C. N. de Araujo}
\email{jcarlos.dearaujo@inpe.br}
\affiliation{Divis\~ao de Astrof\'isica, Instituto Nacional de Pesquisas Espaciais, Avenida dos Astronautas 1758, S\~ao Jos\'e dos Campos, 12227-010, SP, Brazil}

\begin{abstract}
\noindent The Kerr spacetime is a fundamental solution of general relativity (GR), describing the gravitational field around a rotating, uncharged black hole (BH). Kerr spacetime has been crucial in modern astrophysics and it serves as a foundation for the study of gravitational waves (GWs). Possible deviations in Kerr geometry may indicate deviations from GR predictions. In this work, we consider the Johannsen-Psaltis metric, which is a beyond-Kerr metric characterized by a single free parameter, and then we probe this theory framework using several GWs observations from the third Gravitational-wave Transient Catalog (GWTC-3). We find that, for most of the events analyzed, there are no significant deviations from the null hypothesis, i.e. the Kerr metric. Our main findings demonstrate alignment and certain enhancements when compared to previous estimates documented in the literature.
\end{abstract}



\maketitle

\section{Introduction}
\label{sec:introduction}

Since the first direct detection of gravitational waves (GWs) from the incredible observation of GW150914 event \cite{Abbott_2016}, there have been several other new detections from binary mergers, providing ever-increasing data and information on these extreme gravity phenomena. At the current date, almost one hundred coalescing compact binary events have been detected, being of this total, 35 events identified in the third Gravitational-wave Transient Catalog (GWTC-3) \cite{gwtc3}. These GW detections provide us the possibility to perform ever more robust and novel tests of the general relativity (GR) framework in strong field regimes (see \cite{Berti_2015} for a review). 

It is a common understanding that GR is the standard theory of gravity. Not only did it withstand a variety of experimental tests in different settings \cite{Will_2014,Turyshev_2008}, but one may argue that one of its greatest achievements was the prediction of GWs that are being detected today. Nevertheless, it is also a general agreement that GR has its drawbacks and is not a complete theory of gravity, giving rise to the need to take into consideration alternative theories that extend GR beyond its limitations. 

There are theoretical and observational reasons to believe that GR should be modified when gravitational fields are strong and/or on the universe at large scales. From an observational point of view, the physical mechanism responsible for accelerating the Universe at late times is still an open question, and new degrees of freedom of the gravitational origin are alternatives to explain such an accelerated stage (see \cite{Ishak:2018his,Heisenberg_2019,CANTATA:2021ktz} for review). Theories beyond GR can serve as alternatives to explain the current tension in the Hubble constant that persists in the framework of the $\Lambda$CDM model \cite{DiValentino:2021izs}. Also, modified gravity models are motivated to drive the accelerating expansion of the Universe at early times (inflation era). See \cite{Berti_2015} and references therein for motivation of modified gravity (MG) scenarios under the regime of strong gravitational field.

However, concerning tests of GR, the choice of an MG theory against which GR should be tested is another problem entirely. In strong field regimes, this endeavor can be facilitated with the possibility of performing theory-agnostic tests, putting entire sets of alternative metric theories of gravity under scrutiny, through the parameterized post-Einsteinian framework (ppE) \cite{Yunes_2009}. Extra parameters introduced by some MG theories can be mapped to generalized ppE parameters, such that parameter estimation using the ppE framework can establish constraints to specific MG parameters \cite{Yunes_2016}. The ppE framework has been widely and efficiently used for test of GR \cite{Chatziioannou_2012,Barausse_2016,Yunes_2009,Yunes_2016,Carson_2020_a}. Until the present date, regarding GW detection data, no significant evidence for deviations from the GR was found under generic modifications \cite{theligoscientificcollaboration2021tests,Abbott_2021_GR_test}.

To test the very physical nature of space-time metrics is to test GR itself. Efforts in this direction have recently been carried out regarding GR's description of an astrophysical black hole --- the Kerr solution \cite{Roy_Kerr}. In \cite{Gair_2011,Xin_2019, Carson_2020} beyond-Kerr solutions span from the construction of waveforms to be further tested against some forecast bounds in LISA sensitivity up to derivation of corrections to inspiral and ringdown waveforms. Regarding the latter, an approach is outlined in \cite{Hussain_2022} for the computation of deviations to the quasinormal modes of the ringdown phase for a broad class of beyond-GR theories, including beyond-Kerr. Furthermore, a Bayesian analysis was carried in \cite{Dey2022MeasuringDF} with the event signal GW150914 to estimate posterior constraints to beyond-Kerr parameters in the ringdown regime. Deviations from the Kerr geometry have also been tested on electromagnetic signals \cite{Kong_2014,Tripathi_2022,Glampedakis_2021,Liu_2019,Wang_2020,Krawczynski_2018,Cardoso_2016,Bambi_2017,C_rdenas_Avenda_o_2016,Psaltis_2021,V_lkel_2020}. Notably, it has been suggested that GW data can provide more robust constraints on deformation parameters, surpassing the limitations of X-ray instruments, in tests evaluating deviations from the Schwarzschild metric \cite{cardenas2020gravitational} and challenging the Kerr hypothesis \cite{shashank2022bambi}.
    
In this work, our main aim is to search and probe beyond Kerr spacetimes with real data from binary black holes (BBH) coalescence systems from GWTC-3. For this purpose, we will consider the Johannsen-Psaltis (JP) metric \cite{Johannsen_2011}, which is a beyond-Kerr metric, with one single deviation parameter, $\epsilon_3$, that is regular and free of unphysical features, such as closed timelike curves outside of the event horizon or naked singularities. Although it is a simple beyond-GR model, taking into account at most 2 PN corrections, the JP metric also preserves core features of the Kerr solution --- being also axisymmetric, stationary, and asymptotically flat. In addition, the fact that there is only one extra free parameter analyzes several GW events more viable statistically. 

Through Bayesian parameter estimation using data from the GWTC-3 catalog, we find constraints on $\epsilon_3$ to be of $\mathcal{O}(1)$, which is in agreement with \cite{theligoscientificcollaboration2021tests}. Furthermore, our constraints present an improvement over the ones found through x-ray observations of black hole accretion disks, which impose $\epsilon_3 < 5$ \cite{Kong_2014, Bambi_2016}. For most of the events analyzed, we did not find significant deviations of Kerr spacetimes. For certain specific events, we observe a slight statistical trend where $\epsilon_3 > 0$. We explore potential factors contributing to this trend.

This paper is structured as follows: In Section \ref{theory} we present the theoretical model that will be used in our analysis, whose methodology is described in Section \ref{GWTC-3}; in Section \ref{results} we present our results and, in Section \ref{final}, we outline our final considerations and perspectives.

\section{Beyond-Kerr spacetimes and modified gravitational wave}
\label{theory}

In this section, we begin our discussion on the ppE framework introduced in \cite{Yunes_2009}, and how it is used for this paper.  The inspiral phase of the waveform, based on the ppE phenomenological approach, in the frequency domain, is given by 

\begin{equation}
\label{ppE_model}
h(f) = A_{\rm GR}(f)(1+\alpha_{\rm ppE} u^{a_{\rm ppE}}) \exp(i[\Phi_{\rm GR}(f)+\beta_{\rm ppE} u^{b_{\rm ppE}}]),
\end{equation}

where $A_{\rm GR}$ and $\Phi_{\rm GR}$ are the GR predictions for the
amplitude and phase of the waveform, $u = (\pi \mathcal{M} f)^{1/3}$,
$\mathcal{M} = M_t \eta^{3/5}$ is the chirp mass, $M_t = m_1 +m_2$ is the total
mass, $m_1$ and $m_2$ are the masses of the two black holes (BHs),
and $\eta = m_1 m_2/M_t^2$ is the symmetric mass ratio. The various parameters with the “ppE” subscripts describe potential deviations from the GR predictions.

Following \cite{Carson_2020}, we now briefly review how the ppE parameters can be computed for beyond-GR metrics, whose dominant corrections to the inspiral phase can be traced to its $(t,t)$ component. Starting from the Ansatz 

\begin{equation} \label{eq:potentialansatz}
    g_{tt} = -1 + \frac{2M}{r}\left( 1+A\frac{M^p}{r^p} \right) + \mathcal{O}\left( \frac{M^2}{r^2}  \right),
\end{equation}
where the parameters $A$ and $p$ characterize the aforementioned dominant corrections and $M$ is the mass of a BH, one can compute corrections to the  gravitational binding energy,
\begin{equation}\label{eq:bindingE}
    E_b = -\frac{1}{2}\mathcal{M}u^2\left[ 1-\frac{2(2p-1)}{3}Av^{2p} \right],
\end{equation}
and to the GW luminosity,
\begin{equation}\label{eq:gwlum}
    \mathcal{L}_{\rm GW}=\frac{32}{5}\eta^2v^{10}\left[ 1+\frac{4(p+1)}{3}Av^{2p} \right],
\end{equation}
where $v$  is the relative velocity.

Eqs. (\ref{eq:bindingE}) and (\ref{eq:gwlum}) lead to expressions relating the ppE parameters with the general deviation parameters $A$ and $p$ introduced in eq. (\ref{eq:potentialansatz}) (see \cite{Tahura_2018} for details). For the phase ppE parameter, we have 

\begin{equation}
\label{ppE_beta}
\beta_{\rm ppE} = - \frac{15}{16(2p-8)(2p-5)} \gamma_{\dot{f}},
\end{equation}
where 

\begin{equation}
\label{ppE_gamma}
\gamma_{\dot{f}} = \frac{2}{3} \frac{(p+1)(2p+1)}{\eta^{2p/5}}A.
\end{equation}
The $b_{ppE}$  term is given by
\begin{equation}
\label{ppE_b}
b_{\rm ppE} = 2p-5
\end{equation}
Additionally, the amplitude correction is given by

\begin{equation}
\label{ppE_alpha}
\alpha_{\rm ppE} = -\frac{1}{3}(p+1)(2p-1) \frac{A}{\eta^{2p/5}},
\end{equation}
while the $a_{\rm ppE}$ term becomes
\begin{equation}
\label{ppE_a}
a_{\rm ppE} = 2p.
\end{equation}

Eqs. (\ref{ppE_beta})-(\ref{ppE_a}) are fixed once a particular metric is chosen --- in this case, the JP metric, which parameterizes deviations from the Kerr metric. In Boyer-Lindquist coordinates, the latter's metric components are given by
\begin{align*}
    g^{\rm K}_{tt} &= -\left( 1-\frac{2Mr}{\Sigma} \right), \\
    g^{\rm K}_{rr} &= \frac{\Sigma}{\Delta}, \\
    g^{\rm K}_{\theta\theta} &= \Sigma, \\
    g^{\rm K}_{\phi\phi} &= \left( r^2 + a^2 + \frac{2Ma^2r\sin^2\theta}{\Sigma} \right), \\
    g^{\rm K}_{t\phi} &= -\frac{2Mar\sin^2\theta}{\Sigma}, \label{eq:kerrsol}
\end{align*}
where $\Sigma = r^2 + a^2\cos^2\theta$ and $\Delta=r^2-2Mr+a^2$. The deviation from the Kerr solution proposed by the JP metric can be summarized by the deviation function $h(r,\theta)$ \cite{Johannsen_2011}, which is given by
\begin{equation}\label{eq:deviationfunc}
    h(r,\theta) = \sum_{k=0}^\infty \left( \epsilon_{2k} + \epsilon_{2k+1}\frac{Mr}{\Sigma}  \right) \left( \frac{M^2}{\Sigma} \right)^k,
\end{equation}
where $\epsilon_k$ represents the set of deviation parameters. These are constrained by the requirement that the metric be asymptotically flat, which implies $\epsilon_0=\epsilon_1=0$, and by observational constraints obtained through weak-field tests of GR in the parameterized post-Newtonian (PPN) framework \cite{Will_2014}, which implies $|\epsilon_2|\lesssim 10^{-4}$. Constraints for the next leading order parameter, $\epsilon_3$, can be found to be $\epsilon_3 < 4$ through the analysis of the thermal spectrum of geometrically thin BH accretion disks \cite{Kong_2014}. Stronger bounds such as $\epsilon_3\lesssim 10^{-3}$ can also be found through IMR consistency tests simulating data of extreme events expected to be detected by LISA \cite{Carson_2020}. These constraints share the common bound $\epsilon_3>0$, which is found in \cite{Dey2022MeasuringDF} through Bayesian analysis with current and future GW data to test deviations from the multipole structure of the Kerr solution.

\begin{figure*}
    \centering
    \includegraphics[scale=0.4]{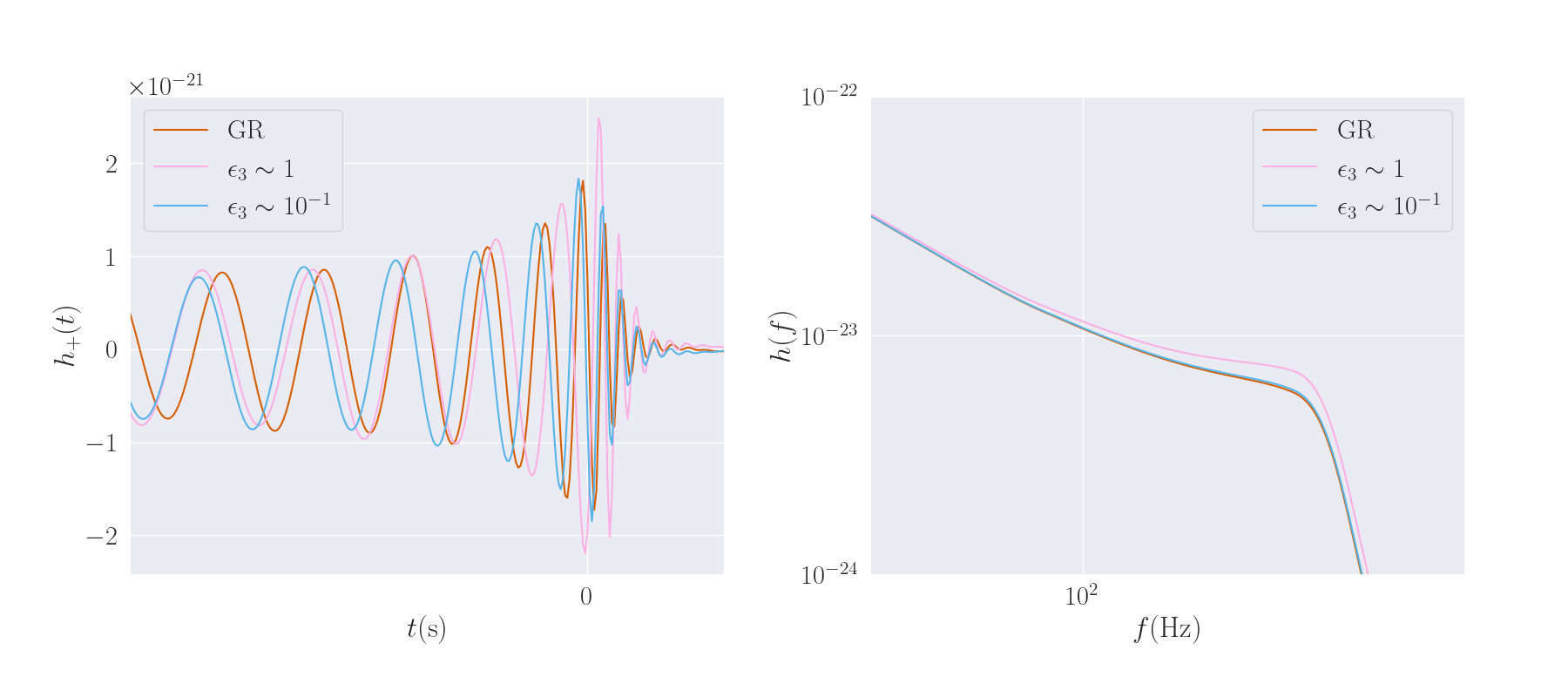}
    \caption{Left panel: GW polarisation signal GW150914-like BH binaries in the detector frame from the GR and ppE model under consideration in this work assuming some deviations from the GR prediction. Right panel: Same as in the left panel, but for the waveform amplitudes in the frequency domain.}
    \label{fig:wavecomparison}
\end{figure*}

Nevertheless, the case considered here is the JP metric such that, except $\epsilon_3$, all other deviation parameters are null. Hence, the deviation function $h(r,\theta)$ given by Eq. (\ref{eq:deviationfunc}) becomes
\begin{equation}
    h(r,\theta)=\epsilon_3\frac{M^3r}{\Sigma^2}.
\end{equation}
Considering this correction, the JP metric is given by
\begin{align*}
    g^{\rm JP}_{tt} &= -\left( 1-\frac{2Mr}{\Sigma} \right) - \epsilon_3\frac{M^3(r-2M)}{r^4}, \\
    g^{\rm JP}_{rr} &= \frac{\Sigma}{\Delta} + \epsilon_3\frac{M^3(r-2M)}{\Delta^2}, \\
    g^{\rm JP}_{\theta\theta} &= \Sigma, \\
    g^{\rm JP}_{\phi\phi} &= \left( r^2 + a^2 + \frac{2Ma^2r\sin^2\theta}{\Sigma} \right) + \epsilon_3\frac{a^2M^3(r+2M)}{r^4}, \\
    g^{\rm JP}_{t\phi} &= -\frac{2Mar\sin^2\theta}{\Sigma} - \epsilon_3\frac{2aM^4}{r^4}. \label{eq:jpmetric}
\end{align*}
Hence, the general deviation parameters can be identified as $p=2$ and $A = - \frac{\epsilon_3}{2}$, yielding the following values for the ppE parameters:
\begin{equation}
\label{JP metric_b}
\beta_{\rm ppE} = \frac{75 \epsilon_3}{64 \eta^{4/5}}, \,\,\,\,\, b_{\rm ppE} = -1,
\end{equation}
\begin{equation}
\label{JP metric_a}
\alpha_{\rm ppE} = \frac{3 \epsilon_3}{2 \eta^{4/5}}, \,\,\,\,\, a_{\rm ppE} = 4;
\end{equation}
the corrections were entered in 2PN order.

Figure \ref{fig:wavecomparison} shows the waveform as predicted by GR together with waveforms from the ppE framework under the beyond-Kerr model. To draw the figure we use the \texttt{pyCBC} \cite{Biwer:2018osg} package with waveform \texttt{IMRPhenomXPHM}. For a quantitative test, we set $\epsilon_3 = 1, 10^{-1}$, and the other 15 baseline parameters to the GW150914 event. We observe a relative phase difference at the order of $\mathcal{O}(10^{-3})$ when comparing the predicted waveforms between the GR and beyond-Kerr frameworks. Additionally, we identify peaks in the amplitude difference occurring at around 250 Hz.

In what follows, we introduce our methodology to be used to probe the full parametric space of the model presented here.

\section{Methodology and GWTC-3}
\label{GWTC-3}

We employ a standard Gaussian noise likelihood $\mathcal{L}$ for the strain data $d$ given the source parameters $\theta$ \cite{Ashton_2019}

\begin{equation}
\label{L}
\ln \mathcal{L}(d|\theta) = -\frac{1}{2} \sum_k \Big\{ \frac{[d_k - h_k(\theta)]^2}{\sigma^2_k} + \ln(2 \pi \sigma^2_k) \Big\},
\end{equation}
where $k$ is the frequency bin index, $\sigma$ is the noise amplitude spectral density, and $h(\theta)$ is the waveform. 

The GW signal emitted from a BBH coalescence depends on intrinsic parameters that directly characterize the 
binary's dynamics and emitted waveform, and extrinsic parameters that encode the relation of the source to 
the detector network. An isolated BH is uniquely described by its mass, spin, and electric charge. Here let us assume that the electric charge is negligible. A BBH undergoing quasi-circular inspiral can be described by eight intrinsic parameters: the primary mass $m_1$, the secondary mass $m_2$ and the three-dimensional spin vectors of the primary spin vector $\vec{S}_1$, and the secondary spin vector $\vec{S}_2$, defined at a reference frequency. Also, we have seven additional extrinsic parameters that are needed to describe a BH binary: the sky location (right ascension $\alpha$ and declination $\delta$); the luminosity distance $d_L$; the orbital inclination $\iota$; the polarization angle $\psi$; the time $t_c$ and phase $\phi_c$ at coalescence. For precessing binaries, the orbital angular momentum is not a stable direction, and it is preferable to describe the source inclination by the angle $\theta_{JN}$ between the total angular momentum and the line-of-sight instead of the orbital inclination angle $\iota$. 

On the other hand, as reviewed in the previous section \ref{theory}, the beyond-Kerr spacetimes to be analyzed in this work are fully characterized by the new and free parameter $\epsilon_3$, which quantifies possible deviations from GR. Therefore, our full parameter space (free parameter baseline) can be defined as 

\begin{eqnarray*}
\theta \equiv\! \Bigl\{m_1, m_2, a_1, a_2, {\rm tilt}_1, {\rm tilt}_2, \phi_{12}, \\
\phi_{JL}, \alpha, \delta, \theta_{JN}, \psi, \phi_c, \epsilon_3 \Bigr\},
\label{eq:parameter_space1}
\end{eqnarray*}
for all BBH mergers. In the definition above, $a_i$ is the dimensionless spin magnitude of the ith object and ${\rm tilt}_i$ is the zenith angle between the spin and orbital angular momenta for the ith object. 

We analyze binary black hole (BBH) coalescence systems from the Gravitational-Wave Transient Catalog 3 (GWTC-3). To conduct this analysis, we utilize the \texttt{PyCBC} Bayesian inference subpackage for gravitational-wave astronomy \cite{Biwer:2018osg}. We have customized this code to integrate our theoretical model as described previously. Additionally, we cross-validate our results for consistency using the \texttt{bilby} code \cite{Ashton_2019}.

To test our main motivation, our event sample consists mainly in: i) We do not consider events where the secondary component, $m_2$, is a neutron star. Thus, we only take into account events where both compact objects are BH. ii) Because we are carrying out a GR test with additional free parameters that can considerably change the predictions on the signal, we only consider events with (signal-to-noise ratio) SNR $>$ 10. This point can be interpreted that we only consider events with very significant signals in their detection.

\begin{table}
	\caption{Constraints on the $\epsilon_3$ parameters at 95\% CL obtained from various BBH coalescence events.}
	\label{tab:ex}
	\begin{tabular}{cccccccc}
		\hline
		Event & $\epsilon_3$\\ 
		\hline
		GW150914 & $0.13^{+0.45}_{-0.27}$\\ 
		GW191129 & $-0.17^{+0.23}_{-0.23}$\\ 
		GW191215 & $0.46^{+1.01}_{-0.61}$\\ 
		GW191222 & $0.42^{+0.48}_{-0.23}$\\ 
		GW191230 & $-0.19^{+0.28}_{-0.28}$\\ 
		GW200112 & $0.13^{+0.61}_{-0.17}$\\ 
		GW200128 & $0.12^{+0.14}_{-0.39}$\\ 
		GW200129 & $-0.01^{+0.32}_{-0.29}$\\ 
		GW200202 & $-0.1^{+0.24}_{-0.13}$\\ 
		GW200208 & $0.22^{+0.75}_{-0.73}$\\ 
		GW200219 & $0.3^{+0.6}_{-0.6}$\\ 
		GW200224 & $0.14^{+0.47}_{-0.47}$\\ 
		GW200302 & $0.13^{+0.56}_{-0.45}$\\ 
		GW200311 & $0.06^{+0.61}_{-0.55}$\\ 
		\hline
	\end{tabular}
\end{table}

\begin{figure*}
    \centering
    \includegraphics[scale=0.7]{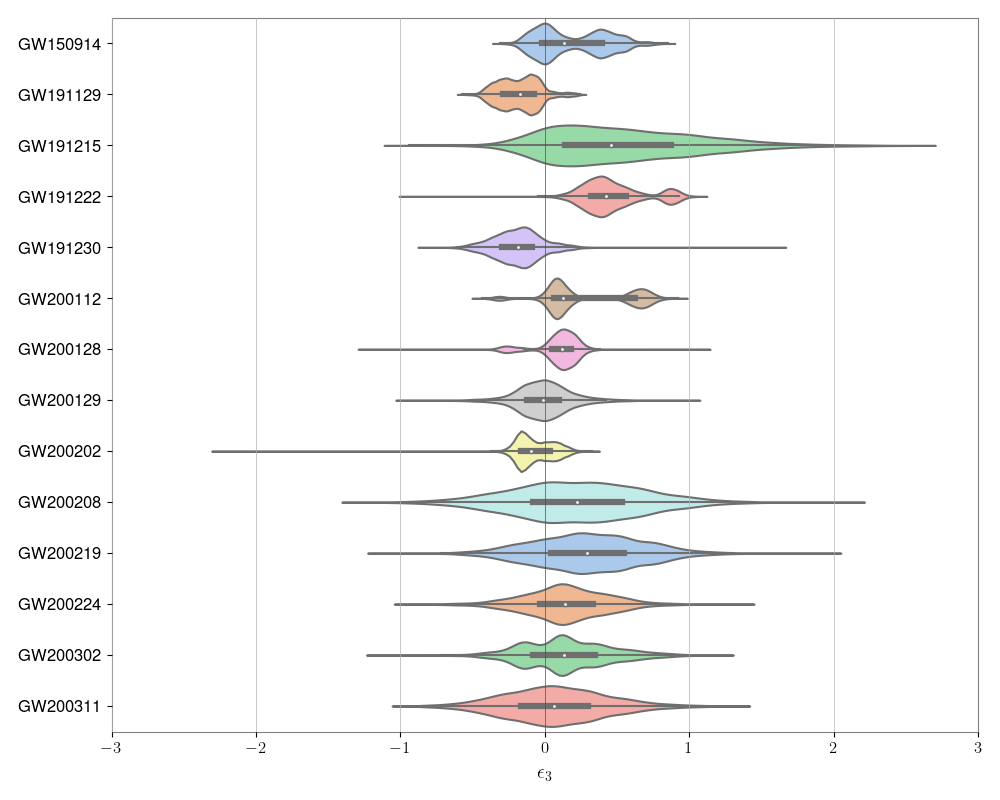}
    \caption{Posterior probability densities for $\epsilon_3$ as recovered for all the events analyzed.The GR value, $\epsilon_3=0$, is represented by the vertical red line.} 
    \label{fig:enter-label}
\end{figure*}

\begin{figure*}
    \centering
    \includegraphics[scale=0.4]{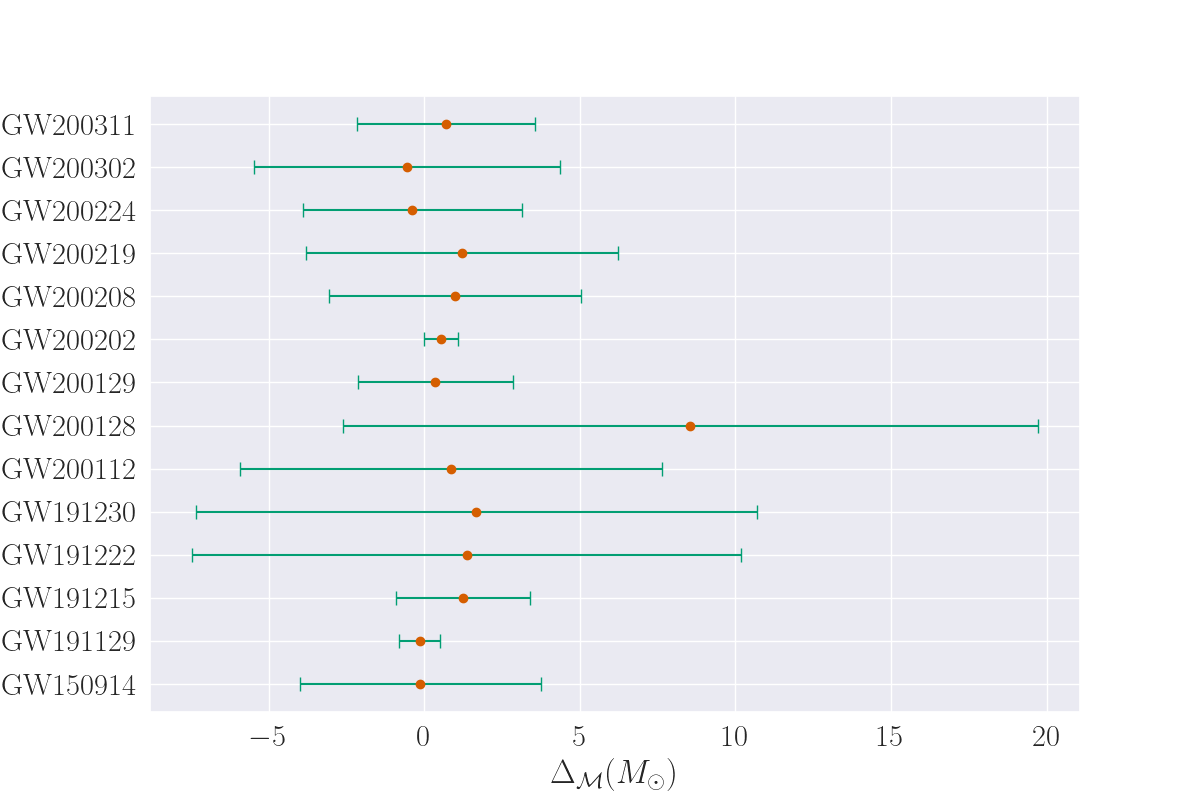}
    \caption{Relative difference at 95\% CL for the chirp mass, $\Delta_\mathcal{M} = \mathcal{M}_{\rm GR} - \mathcal{M}_{ \rm ppE}$, between GR and the model under consideration in this work.}
    \label{fig:diff_chirp}
\end{figure*}

\begin{figure*}
    \centering
    \includegraphics[scale=0.4]{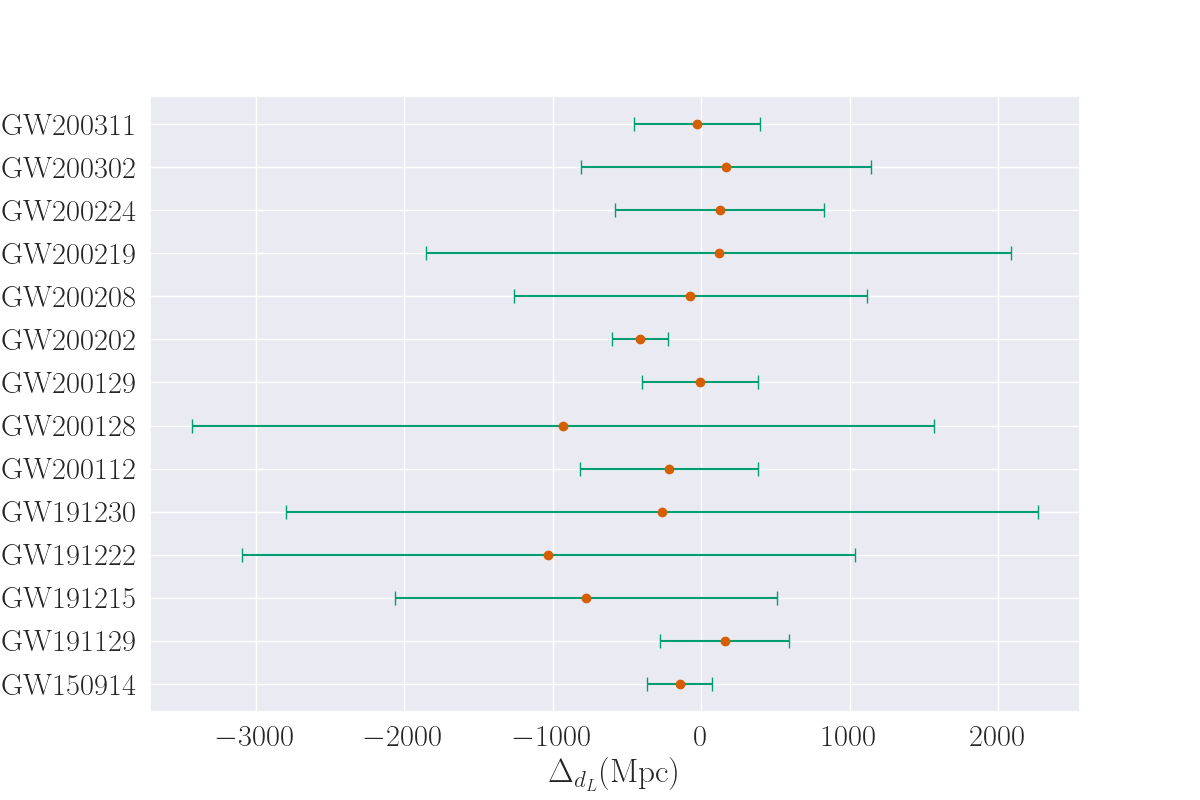}
    \caption{Relative difference at 95\% CL for the luminosity distance $\Delta_{d_L} = d_{L_{\rm GR}} - d_{L_{\rm ppE}}$, between GR and the model under consideration in this work.}
    \label{fig:diff_dL}
\end{figure*}

\section{Main Results}
\label{results}

We analyze 15 events of GWTC-3, using the data available on GWOSC \footnote{https://gwosc.org/}. For our analysis runs, the waveform templates are generated using the \texttt{IMRPhenomXPHM} model, starting at $f_{\rm low}$ = 20 Hz. All free parameters were evaluated in the detector frame. 

In all analyses carried out with phase and amplitude correction, we choose a flat prior $\epsilon_3 \in [-10, 10]$. For all other additional intrinsic and extrinsic parameters of the binary system, we follow the prior range by default corresponding to each event. 
Before analyzing the GWTC-3 events, we first check the model on the GW150914 event. We find $\epsilon_3 = 0.13^{+0.45}_{-0.27}$ at 95\% confidence level (CL). The constraint on $\epsilon_3$ is fully compatible with the null hypothesis, i.e., $\epsilon_3=0$, for greater statistical significance.

Table \ref{tab:ex} summarizes our key results, where we show the constraints at 95\% CL on the parameter $\epsilon_3$ for all events analyzed in this work. We note that the analyzed signals impose constraints of approximately $\mathcal{O}(1)$ magnitude on the beyond-Kerr distortion parameter, considering the expected error bar in interpretation. These results are consistent with the anticipated behavior of the $\phi_4$ parameter, corresponding to corrections at the 2PN order, as presented in \cite{theligoscientificcollaboration2021tests}.
It is noteworthy that these observational constraints align with the findings presented in \cite{Carson_2020}, which adopts a different observational approach (focusing on events within the LISA sensitivity range). Additionally, in \cite{Dey2022MeasuringDF} (see also \cite{ahmed2024constraining}), using solely the ringdown signal of GW150914, the parameter $\epsilon_3$ is constrained at $\mathcal{O}(10)$, demonstrating compatibility with GR and exhibiting a statistical tendency towards positive values rather than negative ranges.

The parameter $\epsilon_3$, constrained by x-ray observations of black hole accretion disks, is bounded by $\epsilon_3 < 5$ \cite{Kong_2014, Bambi_2016}. Consequently, our results indicate a substantial refinement of these constraints when examining GWCT-3 events.

Figure \ref{fig:enter-label} shows the posterior probability densities for $\epsilon_3$ as recovered for all the events analyzed. As mentioned, most events are compatible with $\epsilon_3 =0$. We emphasize that for most of the events analyzed, the parameter $\epsilon_3$ tends to exhibit positive values, yet remains statistically compatible with GR at a high level of significance. This tendency is also noted in \cite{Dey2022MeasuringDF} in the context of ringdown measurements, where it is argued that the region of parameter space $\epsilon_3 > 0$ leads to frequencies and damping times that satisfy their respective bounds for a wide range of remnant masses. We verify that $\epsilon_3$ exhibits no significant correlation with the masses or chirp masses of the events. Consequently, it becomes plausible that during parameter estimation, waveforms with amplitudes exceeding those predicted by GR may receive marginal preference, a scenario also consistent with higher values of $\epsilon_3$, but yet ultimately remains compatible with GR at a high significance level exceeding 2$\sigma$.

To assess changes in key baseline parameters such as masses and distance, we examine the model's ability to approximate GR. In Figures \ref{fig:diff_chirp} and \ref{fig:diff_dL}, we present the disparities in chirp mass and luminosity distance between the GR predictions obtained from the GWOSC database and our analyses based on the ppE framework. Notably, in these figures, subscripts GR and ppE denote values derived from the GWOSC database and our ppE framework analyses, respectively. Consistent with expectations and findings in previous studies, e.g., \cite{theligoscientificcollaboration2021tests}, we observe no significant deviations for the most of events analyzed. In particular, we highlight the event GW200202, characterized by high-mass sources, which potentially renders it relevant in the search for intermediate-mass black hole binaries. This event occurred at a luminosity distance of $d_L = 410^{+150}_{-160}$ Mpc, as predicted by GR. We observe a minor deviation towards $\Delta d_L < 0$ when compared with the predictions of the model considered in this work. This deviation may suggest the presence of some extra cosmological effects on GW200202 if $\Delta d_L < 0$ is not merely a systematical effect/artifact. It is well established that alterations in the gravitational framework can influence predictions of luminosity distance (see \cite{Nishizawa_2018,D_Agostino_2019,Finke_2021} for a few examples). However, in our main baseline parameters, we do not consider this possibility. Thus, we interpret that the $\epsilon_3$ correction on the amplitude of the waveform may systematically mimic some cosmological effect, despite it not being considered. The GW200202 event has also recently been investigated as a dark siren event for measuring the Hubble constant \cite{alfradique2023dark}. This intriguing case warrants further investigation, which we plan to address in future communications.

This conclusion extends to all other physical parameters of the binary systems. While additional parameters beyond GR may modulate signals differently, there are no statistically significant deviations observed between the intrinsic and extrinsic parameters of binary systems under both GR and the ppE framework beyond the Kerr model.

\section{Final remarks}
\label{final}

The detection of GWs not only provided direct evidence for the existence of these waves but also validated various aspects of GR. The observed waveforms have consistently matched the predictions of GR's equations, affirming its accuracy in describing gravitational behavior. Subsequently, observatories like LIGO, Virgo, and KAGRA have continuously detected GWs from diverse sources, ushering in a new era of astronomy and opportunities for additional tests of GR. In this paper, we investigate the JP metric in light of several GW events from the GWTC-3. Our analysis reveals that, for the majority of the scrutinized signals, there is no discernible evidence supporting deviations from a Kerr spacetime. It is important to emphasize that the adjustments made in this study align with the findings outlined in \cite{theligoscientificcollaboration2021tests}. However, it is statistically unrealistic to expect a seamless integration of the parameter $\epsilon_3$ into the extensive phase-only correction analysis described in \cite{theligoscientificcollaboration2021tests}, due to the complexity of the statistical parametric space and the unique functional form of the modified waveform. Nevertheless, we anticipate that the additional free parameters will be of comparable magnitude, although exact replication of statistical outcomes may not be essential. Therefore, within the framework presented here, it is imperative to employ more comprehensive methodologies to thoroughly investigate these signals, offering opportunities for future research.

On the other hand, considerable efforts have been dedicated to probing the nature of spacetime in the presence of strong gravitational fields. Future investigations should extend beyond the specific beyond-Kerr spacetimes considered in this study. Notably, exploring scenarios that incorporate higher-order post-Newtonian (PN) corrections would be valuable. This becomes particularly relevant for forecast analyses targeting upcoming detectors such as the Einstein Telescope and Cosmic Explorer. Additionally, as real data from the fourth Gravitational-Wave Transient Catalog (GWTC-4) becomes available shortly, incorporating these higher-order corrections would provide a more comprehensive understanding.

In forthcoming works, it is essential to develop strategies that involve joint signal analysis in both the electromagnetic and gravitational bands. This approach is crucial for achieving tests with greater consistency in the context of deviations from General Relativity's predictions. The potential integration of data from both domains could offer more robust insights into the behavior of gravitational waves. These avenues for future research represent exciting prospects and will undoubtedly contribute to advancing our understanding of gravitational physics.

\begin{acknowledgments}
\noindent The authors express their gratitude to the referee for providing valuable comments and suggestions aimed at enhancing the overall quality of the work. R.M.S. thanks CNPq (141351/2023-3) for partial financial support. R.C.N thanks the financial support from the Conselho Nacional de Desenvolvimento Cient\'{i}fico e Tecnologico (CNPq, National Council for Scientific and Technological Development) under the project No. 304306/2022-3, and the Fundação de Amparo à pesquisa do Estado do RS (FAPERGS, Research Support Foundation of the State of RS) for partial financial support under the project No. 23/2551-0000848-3. J.C.N.A. thanks FAPESP (2013/26258-4) and CNPq (308367/2019-7) for partial financial support. 
\end{acknowledgments}

\bibliography{PRD}

\begin{thebibliography}{45}%
\makeatletter
\providecommand \@ifxundefined [1]{%
 \@ifx{#1\undefined}
}%
\providecommand \@ifnum [1]{%
 \ifnum #1\expandafter \@firstoftwo
 \else \expandafter \@secondoftwo
 \fi
}%
\providecommand \@ifx [1]{%
 \ifx #1\expandafter \@firstoftwo
 \else \expandafter \@secondoftwo
 \fi
}%
\providecommand \natexlab [1]{#1}%
\providecommand \enquote  [1]{``#1''}%
\providecommand \bibnamefont  [1]{#1}%
\providecommand \bibfnamefont [1]{#1}%
\providecommand \citenamefont [1]{#1}%
\providecommand \href@noop [0]{\@secondoftwo}%
\providecommand \href [0]{\begingroup \@sanitize@url \@href}%
\providecommand \@href[1]{\@@startlink{#1}\@@href}%
\providecommand \@@href[1]{\endgroup#1\@@endlink}%
\providecommand \@sanitize@url [0]{\catcode `\\12\catcode `\$12\catcode
  `\&12\catcode `\#12\catcode `\^12\catcode `\_12\catcode `\%12\relax}%
\providecommand \@@startlink[1]{}%
\providecommand \@@endlink[0]{}%
\providecommand \url  [0]{\begingroup\@sanitize@url \@url }%
\providecommand \@url [1]{\endgroup\@href {#1}{\urlprefix }}%
\providecommand \urlprefix  [0]{URL }%
\providecommand \Eprint [0]{\href }%
\providecommand \doibase [0]{http://dx.doi.org/}%
\providecommand \selectlanguage [0]{\@gobble}%
\providecommand \bibinfo  [0]{\@secondoftwo}%
\providecommand \bibfield  [0]{\@secondoftwo}%
\providecommand \translation [1]{[#1]}%
\providecommand \BibitemOpen [0]{}%
\providecommand \bibitemStop [0]{}%
\providecommand \bibitemNoStop [0]{.\EOS\space}%
\providecommand \EOS [0]{\spacefactor3000\relax}%
\providecommand \BibitemShut  [1]{\csname bibitem#1\endcsname}%
\let\auto@bib@innerbib\@empty
\bibitem [{\citenamefont {Abbott}\ \emph {et~al.}(2016)\citenamefont {Abbott},
  \citenamefont {Abbott}, \citenamefont {Abbott},\ and\ \citenamefont
  {et~al}}]{Abbott_2016}%
  \BibitemOpen
  \bibfield  {author} {\bibinfo {author} {\bibfnamefont {B.}~\bibnamefont
  {Abbott}}, \bibinfo {author} {\bibfnamefont {R.}~\bibnamefont {Abbott}},
  \bibinfo {author} {\bibfnamefont {T.}~\bibnamefont {Abbott}}, \ and\ \bibinfo
  {author} {\bibfnamefont {M.~A.}\ \bibnamefont {et~al}},\ }\href {\doibase
  10.1103/physrevlett.116.241102} {\bibfield  {journal} {\bibinfo  {journal}
  {Physical Review Letters}\ }\textbf {\bibinfo {volume} {116}} (\bibinfo
  {year} {2016}),\ 10.1103/physrevlett.116.241102}\BibitemShut {NoStop}%
\bibitem [{\citenamefont {Collaboration}\ \emph
  {et~al.}(2021{\natexlab{a}})\citenamefont {Collaboration}, \citenamefont {the
  Virgo~Collaboration}, \citenamefont {the KAGRA~Collaboration}, \citenamefont
  {Abbott},\ and\ \citenamefont {et~al}}]{gwtc3}%
  \BibitemOpen
  \bibfield  {author} {\bibinfo {author} {\bibfnamefont {T.~L.~S.}\
  \bibnamefont {Collaboration}}, \bibinfo {author} {\bibnamefont {the
  Virgo~Collaboration}}, \bibinfo {author} {\bibnamefont {the
  KAGRA~Collaboration}}, \bibinfo {author} {\bibfnamefont {R.}~\bibnamefont
  {Abbott}}, \ and\ \bibinfo {author} {\bibfnamefont {T.~D.~A.}\ \bibnamefont
  {et~al}},\ }\href@noop {} {\enquote {\bibinfo {title} {Gwtc-3: Compact binary
  coalescences observed by ligo and virgo during the second part of the third
  observing run},}\ } (\bibinfo {year} {2021}{\natexlab{a}}),\ \Eprint
  {http://arxiv.org/abs/2111.03606} {arXiv:2111.03606 [gr-qc]} \BibitemShut
  {NoStop}%
\bibitem [{\citenamefont {et~al}(2015)}]{Berti_2015}%
  \BibitemOpen
  \bibfield  {author} {\bibinfo {author} {\bibfnamefont {E.~B.}\ \bibnamefont
  {et~al}},\ }\href {\doibase 10.1088/0264-9381/32/24/243001} {\bibfield
  {journal} {\bibinfo  {journal} {Classical and Quantum Gravity}\ }\textbf
  {\bibinfo {volume} {32}},\ \bibinfo {pages} {243001} (\bibinfo {year}
  {2015})}\BibitemShut {NoStop}%
\bibitem [{\citenamefont {Will}(2014)}]{Will_2014}%
  \BibitemOpen
  \bibfield  {author} {\bibinfo {author} {\bibfnamefont {C.~M.}\ \bibnamefont
  {Will}},\ }\href {\doibase 10.12942/lrr-2014-4} {\bibfield  {journal}
  {\bibinfo  {journal} {Living Reviews in Relativity}\ }\textbf {\bibinfo
  {volume} {17}} (\bibinfo {year} {2014}),\ 10.12942/lrr-2014-4}\BibitemShut
  {NoStop}%
\bibitem [{\citenamefont {Turyshev}(2008)}]{Turyshev_2008}%
  \BibitemOpen
  \bibfield  {author} {\bibinfo {author} {\bibfnamefont {S.~G.}\ \bibnamefont
  {Turyshev}},\ }\href@noop {} {\bibfield  {journal} {\bibinfo  {journal}
  {Annual Review of Nuclear and Particle Science}\ }\textbf {\bibinfo {volume}
  {58}},\ \bibinfo {pages} {207} (\bibinfo {year} {2008})}\BibitemShut
  {NoStop}%
\bibitem [{\citenamefont {Ishak}(2019)}]{Ishak:2018his}%
  \BibitemOpen
  \bibfield  {author} {\bibinfo {author} {\bibfnamefont {M.}~\bibnamefont
  {Ishak}},\ }\href {\doibase 10.1007/s41114-018-0017-4} {\bibfield  {journal}
  {\bibinfo  {journal} {Living Rev. Rel.}\ }\textbf {\bibinfo {volume} {22}},\
  \bibinfo {pages} {1} (\bibinfo {year} {2019})},\ \Eprint
  {http://arxiv.org/abs/1806.10122} {arXiv:1806.10122 [astro-ph.CO]}
  \BibitemShut {NoStop}%
\bibitem [{\citenamefont {Heisenberg}(2019)}]{Heisenberg_2019}%
  \BibitemOpen
  \bibfield  {author} {\bibinfo {author} {\bibfnamefont {L.}~\bibnamefont
  {Heisenberg}},\ }\href {\doibase 10.1016/j.physrep.2018.11.006} {\bibfield
  {journal} {\bibinfo  {journal} {Physics Reports}\ }\textbf {\bibinfo {volume}
  {796}},\ \bibinfo {pages} {1} (\bibinfo {year} {2019})}\BibitemShut {NoStop}%
\bibitem [{\citenamefont {Saridakis}\ \emph {et~al.}(2021)\citenamefont
  {Saridakis} \emph {et~al.}}]{CANTATA:2021ktz}%
  \BibitemOpen
  \bibfield  {author} {\bibinfo {author} {\bibfnamefont {E.~N.}\ \bibnamefont
  {Saridakis}} \emph {et~al.} (\bibinfo {collaboration} {CANTATA}),\
  }\href@noop {} {\  (\bibinfo {year} {2021})},\ \Eprint
  {http://arxiv.org/abs/2105.12582} {arXiv:2105.12582 [gr-qc]} \BibitemShut
  {NoStop}%
\bibitem [{\citenamefont {Di~Valentino}\ \emph {et~al.}(2021)\citenamefont
  {Di~Valentino}, \citenamefont {Mena}, \citenamefont {Pan}, \citenamefont
  {Visinelli}, \citenamefont {Yang}, \citenamefont {Melchiorri}, \citenamefont
  {Mota}, \citenamefont {Riess},\ and\ \citenamefont
  {Silk}}]{DiValentino:2021izs}%
  \BibitemOpen
  \bibfield  {author} {\bibinfo {author} {\bibfnamefont {E.}~\bibnamefont
  {Di~Valentino}}, \bibinfo {author} {\bibfnamefont {O.}~\bibnamefont {Mena}},
  \bibinfo {author} {\bibfnamefont {S.}~\bibnamefont {Pan}}, \bibinfo {author}
  {\bibfnamefont {L.}~\bibnamefont {Visinelli}}, \bibinfo {author}
  {\bibfnamefont {W.}~\bibnamefont {Yang}}, \bibinfo {author} {\bibfnamefont
  {A.}~\bibnamefont {Melchiorri}}, \bibinfo {author} {\bibfnamefont {D.~F.}\
  \bibnamefont {Mota}}, \bibinfo {author} {\bibfnamefont {A.~G.}\ \bibnamefont
  {Riess}}, \ and\ \bibinfo {author} {\bibfnamefont {J.}~\bibnamefont {Silk}},\
  }\href {\doibase 10.1088/1361-6382/ac086d} {\bibfield  {journal} {\bibinfo
  {journal} {Class. Quant. Grav.}\ }\textbf {\bibinfo {volume} {38}},\ \bibinfo
  {pages} {153001} (\bibinfo {year} {2021})},\ \Eprint
  {http://arxiv.org/abs/2103.01183} {arXiv:2103.01183 [astro-ph.CO]}
  \BibitemShut {NoStop}%
\bibitem [{\citenamefont {Yunes}\ and\ \citenamefont
  {Pretorius}(2009)}]{Yunes_2009}%
  \BibitemOpen
  \bibfield  {author} {\bibinfo {author} {\bibfnamefont {N.}~\bibnamefont
  {Yunes}}\ and\ \bibinfo {author} {\bibfnamefont {F.}~\bibnamefont
  {Pretorius}},\ }\href {\doibase 10.1103/physrevd.80.122003} {\bibfield
  {journal} {\bibinfo  {journal} {Physical Review D}\ }\textbf {\bibinfo
  {volume} {80}} (\bibinfo {year} {2009}),\
  10.1103/physrevd.80.122003}\BibitemShut {NoStop}%
\bibitem [{\citenamefont {Yunes}\ \emph {et~al.}(2016)\citenamefont {Yunes},
  \citenamefont {Yagi},\ and\ \citenamefont {Pretorius}}]{Yunes_2016}%
  \BibitemOpen
  \bibfield  {author} {\bibinfo {author} {\bibfnamefont {N.}~\bibnamefont
  {Yunes}}, \bibinfo {author} {\bibfnamefont {K.}~\bibnamefont {Yagi}}, \ and\
  \bibinfo {author} {\bibfnamefont {F.}~\bibnamefont {Pretorius}},\ }\href@noop
  {} {\bibfield  {journal} {\bibinfo  {journal} {Physical review D}\ }\textbf
  {\bibinfo {volume} {94}},\ \bibinfo {pages} {084002} (\bibinfo {year}
  {2016})}\BibitemShut {NoStop}%
\bibitem [{\citenamefont {Chatziioannou}\ \emph {et~al.}(2012)\citenamefont
  {Chatziioannou}, \citenamefont {Yunes},\ and\ \citenamefont
  {Cornish}}]{Chatziioannou_2012}%
  \BibitemOpen
  \bibfield  {author} {\bibinfo {author} {\bibfnamefont {K.}~\bibnamefont
  {Chatziioannou}}, \bibinfo {author} {\bibfnamefont {N.}~\bibnamefont
  {Yunes}}, \ and\ \bibinfo {author} {\bibfnamefont {N.}~\bibnamefont
  {Cornish}},\ }\href {\doibase 10.1103/physrevd.86.022004} {\bibfield
  {journal} {\bibinfo  {journal} {Physical Review D}\ }\textbf {\bibinfo
  {volume} {86}} (\bibinfo {year} {2012}),\
  10.1103/physrevd.86.022004}\BibitemShut {NoStop}%
\bibitem [{\citenamefont {Barausse}\ \emph {et~al.}(2016)\citenamefont
  {Barausse}, \citenamefont {Yunes},\ and\ \citenamefont
  {Chamberlain}}]{Barausse_2016}%
  \BibitemOpen
  \bibfield  {author} {\bibinfo {author} {\bibfnamefont {E.}~\bibnamefont
  {Barausse}}, \bibinfo {author} {\bibfnamefont {N.}~\bibnamefont {Yunes}}, \
  and\ \bibinfo {author} {\bibfnamefont {K.}~\bibnamefont {Chamberlain}},\
  }\href {\doibase 10.1103/physrevlett.116.241104} {\bibfield  {journal}
  {\bibinfo  {journal} {Physical Review Letters}\ }\textbf {\bibinfo {volume}
  {116}} (\bibinfo {year} {2016}),\ 10.1103/physrevlett.116.241104}\BibitemShut
  {NoStop}%
\bibitem [{\citenamefont {Carson}\ and\ \citenamefont
  {Yagi}(2020{\natexlab{a}})}]{Carson_2020_a}%
  \BibitemOpen
  \bibfield  {author} {\bibinfo {author} {\bibfnamefont {Z.}~\bibnamefont
  {Carson}}\ and\ \bibinfo {author} {\bibfnamefont {K.}~\bibnamefont {Yagi}},\
  }\href {\doibase 10.1103/physrevd.101.044047} {\bibfield  {journal} {\bibinfo
   {journal} {Physical Review D}\ }\textbf {\bibinfo {volume} {101}} (\bibinfo
  {year} {2020}{\natexlab{a}}),\ 10.1103/physrevd.101.044047}\BibitemShut
  {NoStop}%
\bibitem [{\citenamefont {Collaboration}\ \emph
  {et~al.}(2021{\natexlab{b}})\citenamefont {Collaboration}, \citenamefont {the
  Virgo~Collaboration},\ and\ \citenamefont {the
  KAGRA~Collaboration}}]{theligoscientificcollaboration2021tests}%
  \BibitemOpen
  \bibfield  {author} {\bibinfo {author} {\bibfnamefont {T.~L.~S.}\
  \bibnamefont {Collaboration}}, \bibinfo {author} {\bibnamefont {the
  Virgo~Collaboration}}, \ and\ \bibinfo {author} {\bibnamefont {the
  KAGRA~Collaboration}},\ }\href@noop {} {\enquote {\bibinfo {title} {Tests of
  general relativity with gwtc-3},}\ } (\bibinfo {year} {2021}{\natexlab{b}}),\
  \Eprint {http://arxiv.org/abs/2112.06861} {arXiv:2112.06861 [gr-qc]}
  \BibitemShut {NoStop}%
\bibitem [{\citenamefont {et~al}(2021)}]{Abbott_2021_GR_test}%
  \BibitemOpen
  \bibfield  {author} {\bibinfo {author} {\bibfnamefont {R.~A.}\ \bibnamefont
  {et~al}},\ }\href {\doibase 10.1103/physrevd.103.122002} {\bibfield
  {journal} {\bibinfo  {journal} {Physical Review D}\ }\textbf {\bibinfo
  {volume} {103}} (\bibinfo {year} {2021}),\
  10.1103/physrevd.103.122002}\BibitemShut {NoStop}%
\bibitem [{\citenamefont {Kerr}(1963)}]{Roy_Kerr}%
  \BibitemOpen
  \bibfield  {author} {\bibinfo {author} {\bibfnamefont {R.~P.}\ \bibnamefont
  {Kerr}},\ }\href {\doibase 10.1103/PhysRevLett.11.237} {\bibfield  {journal}
  {\bibinfo  {journal} {Phys. Rev. Lett.}\ }\textbf {\bibinfo {volume} {11}},\
  \bibinfo {pages} {237} (\bibinfo {year} {1963})}\BibitemShut {NoStop}%
\bibitem [{\citenamefont {Gair}\ and\ \citenamefont {Yunes}(2011)}]{Gair_2011}%
  \BibitemOpen
  \bibfield  {author} {\bibinfo {author} {\bibfnamefont {J.}~\bibnamefont
  {Gair}}\ and\ \bibinfo {author} {\bibfnamefont {N.}~\bibnamefont {Yunes}},\
  }\href {\doibase 10.1103/physrevd.84.064016} {\bibfield  {journal} {\bibinfo
  {journal} {Physical Review D}\ }\textbf {\bibinfo {volume} {84}} (\bibinfo
  {year} {2011}),\ 10.1103/physrevd.84.064016}\BibitemShut {NoStop}%
\bibitem [{\citenamefont {Xin}\ \emph {et~al.}(2019)\citenamefont {Xin},
  \citenamefont {Han},\ and\ \citenamefont {Yang}}]{Xin_2019}%
  \BibitemOpen
  \bibfield  {author} {\bibinfo {author} {\bibfnamefont {S.}~\bibnamefont
  {Xin}}, \bibinfo {author} {\bibfnamefont {W.-B.}\ \bibnamefont {Han}}, \ and\
  \bibinfo {author} {\bibfnamefont {S.-C.}\ \bibnamefont {Yang}},\ }\href
  {\doibase 10.1103/physrevd.100.084055} {\bibfield  {journal} {\bibinfo
  {journal} {Physical Review D}\ }\textbf {\bibinfo {volume} {100}} (\bibinfo
  {year} {2019}),\ 10.1103/physrevd.100.084055}\BibitemShut {NoStop}%
\bibitem [{\citenamefont {Carson}\ and\ \citenamefont
  {Yagi}(2020{\natexlab{b}})}]{Carson_2020}%
  \BibitemOpen
  \bibfield  {author} {\bibinfo {author} {\bibfnamefont {Z.}~\bibnamefont
  {Carson}}\ and\ \bibinfo {author} {\bibfnamefont {K.}~\bibnamefont {Yagi}},\
  }\href {\doibase 10.1103/physrevd.101.084050} {\bibfield  {journal} {\bibinfo
   {journal} {arXiv: General Relativity and Quantum Cosmology}\ } (\bibinfo
  {year} {2020}{\natexlab{b}}),\ 10.1103/physrevd.101.084050}\BibitemShut
  {NoStop}%
\bibitem [{\citenamefont {Hussain}\ and\ \citenamefont
  {Zimmerman}(2022)}]{Hussain_2022}%
  \BibitemOpen
  \bibfield  {author} {\bibinfo {author} {\bibfnamefont {A.}~\bibnamefont
  {Hussain}}\ and\ \bibinfo {author} {\bibfnamefont {A.}~\bibnamefont
  {Zimmerman}},\ }\href {\doibase 10.1103/physrevd.106.104018} {\bibfield
  {journal} {\bibinfo  {journal} {Physical Review D}\ }\textbf {\bibinfo
  {volume} {106}} (\bibinfo {year} {2022}),\
  10.1103/physrevd.106.104018}\BibitemShut {NoStop}%
\bibitem [{\citenamefont {Dey}\ \emph {et~al.}(2022)\citenamefont {Dey},
  \citenamefont {Barausse},\ and\ \citenamefont {Basak}}]{Dey2022MeasuringDF}%
  \BibitemOpen
  \bibfield  {author} {\bibinfo {author} {\bibfnamefont {K.}~\bibnamefont
  {Dey}}, \bibinfo {author} {\bibfnamefont {E.}~\bibnamefont {Barausse}}, \
  and\ \bibinfo {author} {\bibfnamefont {S.}~\bibnamefont {Basak}},\ }\href
  {https://api.semanticscholar.org/CorpusID:254926592} {\bibfield  {journal}
  {\bibinfo  {journal} {Physical Review D}\ } (\bibinfo {year}
  {2022})}\BibitemShut {NoStop}%
\bibitem [{\citenamefont {Kong}\ \emph {et~al.}(2014)\citenamefont {Kong},
  \citenamefont {Li},\ and\ \citenamefont {Bambi}}]{Kong_2014}%
  \BibitemOpen
  \bibfield  {author} {\bibinfo {author} {\bibfnamefont {L.}~\bibnamefont
  {Kong}}, \bibinfo {author} {\bibfnamefont {Z.}~\bibnamefont {Li}}, \ and\
  \bibinfo {author} {\bibfnamefont {C.}~\bibnamefont {Bambi}},\ }\href
  {\doibase 10.1088/0004-637x/797/2/78} {\bibfield  {journal} {\bibinfo
  {journal} {The Astrophysical Journal}\ }\textbf {\bibinfo {volume} {797}},\
  \bibinfo {pages} {78} (\bibinfo {year} {2014})}\BibitemShut {NoStop}%
\bibitem [{\citenamefont {Tripathi}\ \emph {et~al.}(2022)\citenamefont
  {Tripathi}, \citenamefont {Abdikamalov}, \citenamefont {Ayzenberg},
  \citenamefont {Bambi}, \citenamefont {Grinberg}, \citenamefont {Liu},\ and\
  \citenamefont {Zhou}}]{Tripathi_2022}%
  \BibitemOpen
  \bibfield  {author} {\bibinfo {author} {\bibfnamefont {A.}~\bibnamefont
  {Tripathi}}, \bibinfo {author} {\bibfnamefont {A.~B.}\ \bibnamefont
  {Abdikamalov}}, \bibinfo {author} {\bibfnamefont {D.}~\bibnamefont
  {Ayzenberg}}, \bibinfo {author} {\bibfnamefont {C.}~\bibnamefont {Bambi}},
  \bibinfo {author} {\bibfnamefont {V.}~\bibnamefont {Grinberg}}, \bibinfo
  {author} {\bibfnamefont {H.}~\bibnamefont {Liu}}, \ and\ \bibinfo {author}
  {\bibfnamefont {M.}~\bibnamefont {Zhou}},\ }\href {\doibase
  10.1088/1475-7516/2022/01/019} {\bibfield  {journal} {\bibinfo  {journal}
  {Journal of Cosmology and Astroparticle Physics}\ }\textbf {\bibinfo {volume}
  {2022}},\ \bibinfo {pages} {019} (\bibinfo {year} {2022})}\BibitemShut
  {NoStop}%
\bibitem [{\citenamefont {Glampedakis}\ and\ \citenamefont
  {Pappas}(2021)}]{Glampedakis_2021}%
  \BibitemOpen
  \bibfield  {author} {\bibinfo {author} {\bibfnamefont {K.}~\bibnamefont
  {Glampedakis}}\ and\ \bibinfo {author} {\bibfnamefont {G.}~\bibnamefont
  {Pappas}},\ }\href {\doibase 10.1103/physrevd.104.l081503} {\bibfield
  {journal} {\bibinfo  {journal} {Physical Review D}\ }\textbf {\bibinfo
  {volume} {104}} (\bibinfo {year} {2021}),\
  10.1103/physrevd.104.l081503}\BibitemShut {NoStop}%
\bibitem [{\citenamefont {Liu}\ \emph {et~al.}(2019)\citenamefont {Liu},
  \citenamefont {Abdikamalov}, \citenamefont {Ayzenberg}, \citenamefont
  {Bambi}, \citenamefont {Dauser}, \citenamefont {Garc{\'{\i} }a},\ and\
  \citenamefont {Nampalliwar}}]{Liu_2019}%
  \BibitemOpen
  \bibfield  {author} {\bibinfo {author} {\bibfnamefont {H.}~\bibnamefont
  {Liu}}, \bibinfo {author} {\bibfnamefont {A.~B.}\ \bibnamefont
  {Abdikamalov}}, \bibinfo {author} {\bibfnamefont {D.}~\bibnamefont
  {Ayzenberg}}, \bibinfo {author} {\bibfnamefont {C.}~\bibnamefont {Bambi}},
  \bibinfo {author} {\bibfnamefont {T.}~\bibnamefont {Dauser}}, \bibinfo
  {author} {\bibfnamefont {J.~A.}\ \bibnamefont {Garc{\'{\i} }a}}, \ and\
  \bibinfo {author} {\bibfnamefont {S.}~\bibnamefont {Nampalliwar}},\ }\href
  {\doibase 10.1103/physrevd.99.123007} {\bibfield  {journal} {\bibinfo
  {journal} {Physical Review D}\ }\textbf {\bibinfo {volume} {99}} (\bibinfo
  {year} {2019}),\ 10.1103/physrevd.99.123007}\BibitemShut {NoStop}%
\bibitem [{\citenamefont {Wang}\ \emph {et~al.}(2020)\citenamefont {Wang},
  \citenamefont {Abdikamalov}, \citenamefont {Ayzenberg}, \citenamefont
  {Bambi}, \citenamefont {Dauser}, \citenamefont {Garc{\'{\i} }a},
  \citenamefont {Nampalliwar},\ and\ \citenamefont {Steiner}}]{Wang_2020}%
  \BibitemOpen
  \bibfield  {author} {\bibinfo {author} {\bibfnamefont {J.}~\bibnamefont
  {Wang}}, \bibinfo {author} {\bibfnamefont {A.~B.}\ \bibnamefont
  {Abdikamalov}}, \bibinfo {author} {\bibfnamefont {D.}~\bibnamefont
  {Ayzenberg}}, \bibinfo {author} {\bibfnamefont {C.}~\bibnamefont {Bambi}},
  \bibinfo {author} {\bibfnamefont {T.}~\bibnamefont {Dauser}}, \bibinfo
  {author} {\bibfnamefont {J.~A.}\ \bibnamefont {Garc{\'{\i} }a}}, \bibinfo
  {author} {\bibfnamefont {S.}~\bibnamefont {Nampalliwar}}, \ and\ \bibinfo
  {author} {\bibfnamefont {J.~F.}\ \bibnamefont {Steiner}},\ }\href {\doibase
  10.1088/1475-7516/2020/05/026} {\bibfield  {journal} {\bibinfo  {journal}
  {Journal of Cosmology and Astroparticle Physics}\ }\textbf {\bibinfo {volume}
  {2020}},\ \bibinfo {pages} {026} (\bibinfo {year} {2020})}\BibitemShut
  {NoStop}%
\bibitem [{\citenamefont {Krawczynski}(2018)}]{Krawczynski_2018}%
  \BibitemOpen
  \bibfield  {author} {\bibinfo {author} {\bibfnamefont {H.}~\bibnamefont
  {Krawczynski}},\ }\href {\doibase 10.1007/s10714-018-2419-8} {\bibfield
  {journal} {\bibinfo  {journal} {General Relativity and Gravitation}\ }\textbf
  {\bibinfo {volume} {50}} (\bibinfo {year} {2018}),\
  10.1007/s10714-018-2419-8}\BibitemShut {NoStop}%
\bibitem [{\citenamefont {Cardoso}\ and\ \citenamefont
  {Gualtieri}(2016)}]{Cardoso_2016}%
  \BibitemOpen
  \bibfield  {author} {\bibinfo {author} {\bibfnamefont {V.}~\bibnamefont
  {Cardoso}}\ and\ \bibinfo {author} {\bibfnamefont {L.}~\bibnamefont
  {Gualtieri}},\ }\href {\doibase 10.1088/0264-9381/33/17/174001} {\bibfield
  {journal} {\bibinfo  {journal} {Classical and Quantum Gravity}\ }\textbf
  {\bibinfo {volume} {33}},\ \bibinfo {pages} {174001} (\bibinfo {year}
  {2016})}\BibitemShut {NoStop}%
\bibitem [{\citenamefont {Bambi}\ \emph {et~al.}(2017)\citenamefont {Bambi},
  \citenamefont {C{\'{a} }rdenas-Avenda{\~{n}}o}, \citenamefont {Dauser},
  \citenamefont {Garc{\'{\i}}a},\ and\ \citenamefont
  {Nampalliwar}}]{Bambi_2017}%
  \BibitemOpen
  \bibfield  {author} {\bibinfo {author} {\bibfnamefont {C.}~\bibnamefont
  {Bambi}}, \bibinfo {author} {\bibfnamefont {A.}~\bibnamefont {C{\'{a}
  }rdenas-Avenda{\~{n}}o}}, \bibinfo {author} {\bibfnamefont {T.}~\bibnamefont
  {Dauser}}, \bibinfo {author} {\bibfnamefont {J.~A.}\ \bibnamefont
  {Garc{\'{\i}}a}}, \ and\ \bibinfo {author} {\bibfnamefont {S.}~\bibnamefont
  {Nampalliwar}},\ }\href {\doibase 10.3847/1538-4357/aa74c0} {\bibfield
  {journal} {\bibinfo  {journal} {The Astrophysical Journal}\ }\textbf
  {\bibinfo {volume} {842}},\ \bibinfo {pages} {76} (\bibinfo {year}
  {2017})}\BibitemShut {NoStop}%
\bibitem [{\citenamefont {C{\'{a}}rdenas-Avenda{\~{n}}o}\ \emph
  {et~al.}(2016)\citenamefont {C{\'{a}}rdenas-Avenda{\~{n}}o}, \citenamefont
  {Jiang},\ and\ \citenamefont {Bambi}}]{C_rdenas_Avenda_o_2016}%
  \BibitemOpen
  \bibfield  {author} {\bibinfo {author} {\bibfnamefont {A.}~\bibnamefont
  {C{\'{a}}rdenas-Avenda{\~{n}}o}}, \bibinfo {author} {\bibfnamefont
  {J.}~\bibnamefont {Jiang}}, \ and\ \bibinfo {author} {\bibfnamefont
  {C.}~\bibnamefont {Bambi}},\ }\href {\doibase 10.1016/j.physletb.2016.06.075}
  {\bibfield  {journal} {\bibinfo  {journal} {Physics Letters B}\ }\textbf
  {\bibinfo {volume} {760}},\ \bibinfo {pages} {254} (\bibinfo {year}
  {2016})}\BibitemShut {NoStop}%
\bibitem [{\citenamefont {Psaltis}\ \emph {et~al.}(2021)\citenamefont
  {Psaltis}, \citenamefont {Talbot}, \citenamefont {Payne},\ and\ \citenamefont
  {Mandel}}]{Psaltis_2021}%
  \BibitemOpen
  \bibfield  {author} {\bibinfo {author} {\bibfnamefont {D.}~\bibnamefont
  {Psaltis}}, \bibinfo {author} {\bibfnamefont {C.}~\bibnamefont {Talbot}},
  \bibinfo {author} {\bibfnamefont {E.}~\bibnamefont {Payne}}, \ and\ \bibinfo
  {author} {\bibfnamefont {I.}~\bibnamefont {Mandel}},\ }\href {\doibase
  10.1103/physrevd.103.104036} {\bibfield  {journal} {\bibinfo  {journal}
  {Physical Review D}\ }\textbf {\bibinfo {volume} {103}} (\bibinfo {year}
  {2021}),\ 10.1103/physrevd.103.104036}\BibitemShut {NoStop}%
\bibitem [{\citenamefont {Völkel}\ and\ \citenamefont
  {Barausse}(2020)}]{V_lkel_2020}%
  \BibitemOpen
  \bibfield  {author} {\bibinfo {author} {\bibfnamefont {S.~H.}\ \bibnamefont
  {Völkel}}\ and\ \bibinfo {author} {\bibfnamefont {E.}~\bibnamefont
  {Barausse}},\ }\href {\doibase 10.1103/physrevd.102.084025} {\bibfield
  {journal} {\bibinfo  {journal} {Physical Review D}\ }\textbf {\bibinfo
  {volume} {102}} (\bibinfo {year} {2020}),\
  10.1103/physrevd.102.084025}\BibitemShut {NoStop}%
\bibitem [{\citenamefont {Cardenas-Avendano}\ \emph {et~al.}(2020)\citenamefont
  {Cardenas-Avendano}, \citenamefont {Nampalliwar},\ and\ \citenamefont
  {Yunes}}]{cardenas2020gravitational}%
  \BibitemOpen
  \bibfield  {author} {\bibinfo {author} {\bibfnamefont {A.}~\bibnamefont
  {Cardenas-Avendano}}, \bibinfo {author} {\bibfnamefont {S.}~\bibnamefont
  {Nampalliwar}}, \ and\ \bibinfo {author} {\bibfnamefont {N.}~\bibnamefont
  {Yunes}},\ }\href@noop {} {\bibfield  {journal} {\bibinfo  {journal}
  {Classical and Quantum Gravity}\ }\textbf {\bibinfo {volume} {37}},\ \bibinfo
  {pages} {135008} (\bibinfo {year} {2020})}\BibitemShut {NoStop}%
\bibitem [{\citenamefont {Shashank}\ and\ \citenamefont
  {Bambi}(2022)}]{shashank2022bambi}%
  \BibitemOpen
  \bibfield  {author} {\bibinfo {author} {\bibfnamefont {S.}~\bibnamefont
  {Shashank}}\ and\ \bibinfo {author} {\bibfnamefont {C.}~\bibnamefont
  {Bambi}},\ }\href {\doibase 10.1103/PhysRevD.105.104004} {\bibfield
  {journal} {\bibinfo  {journal} {Phys. Rev. D}\ }\textbf {\bibinfo {volume}
  {105}},\ \bibinfo {pages} {104004} (\bibinfo {year} {2022})}\BibitemShut
  {NoStop}%
\bibitem [{\citenamefont {Johannsen}\ and\ \citenamefont
  {Psaltis}(2011)}]{Johannsen_2011}%
  \BibitemOpen
  \bibfield  {author} {\bibinfo {author} {\bibfnamefont {T.}~\bibnamefont
  {Johannsen}}\ and\ \bibinfo {author} {\bibfnamefont {D.}~\bibnamefont
  {Psaltis}},\ }\href {\doibase 10.1103/physrevd.83.124015} {\bibfield
  {journal} {\bibinfo  {journal} {Physical Review D}\ }\textbf {\bibinfo
  {volume} {83}} (\bibinfo {year} {2011}),\
  10.1103/physrevd.83.124015}\BibitemShut {NoStop}%
\bibitem [{\citenamefont {Bambi}\ \emph {et~al.}(2016)\citenamefont {Bambi},
  \citenamefont {Jiang},\ and\ \citenamefont {Steiner}}]{Bambi_2016}%
  \BibitemOpen
  \bibfield  {author} {\bibinfo {author} {\bibfnamefont {C.}~\bibnamefont
  {Bambi}}, \bibinfo {author} {\bibfnamefont {J.}~\bibnamefont {Jiang}}, \ and\
  \bibinfo {author} {\bibfnamefont {J.~F.}\ \bibnamefont {Steiner}},\ }\href
  {\doibase 10.1088/0264-9381/33/6/064001} {\bibfield  {journal} {\bibinfo
  {journal} {Classical and Quantum Gravity}\ }\textbf {\bibinfo {volume}
  {33}},\ \bibinfo {pages} {064001} (\bibinfo {year} {2016})}\BibitemShut
  {NoStop}%
\bibitem [{\citenamefont {Tahura}\ and\ \citenamefont
  {Yagi}(2018)}]{Tahura_2018}%
  \BibitemOpen
  \bibfield  {author} {\bibinfo {author} {\bibfnamefont {S.}~\bibnamefont
  {Tahura}}\ and\ \bibinfo {author} {\bibfnamefont {K.}~\bibnamefont {Yagi}},\
  }\href {\doibase 10.1103/physrevd.98.084042} {\bibfield  {journal} {\bibinfo
  {journal} {Physical Review D}\ }\textbf {\bibinfo {volume} {98}} (\bibinfo
  {year} {2018}),\ 10.1103/physrevd.98.084042}\BibitemShut {NoStop}%
\bibitem [{\citenamefont {Biwer}\ \emph {et~al.}(2019)\citenamefont {Biwer},
  \citenamefont {Capano}, \citenamefont {De}, \citenamefont {Cabero},
  \citenamefont {Brown}, \citenamefont {Nitz},\ and\ \citenamefont
  {Raymond}}]{Biwer:2018osg}%
  \BibitemOpen
  \bibfield  {author} {\bibinfo {author} {\bibfnamefont {C.~M.}\ \bibnamefont
  {Biwer}}, \bibinfo {author} {\bibfnamefont {C.~D.}\ \bibnamefont {Capano}},
  \bibinfo {author} {\bibfnamefont {S.}~\bibnamefont {De}}, \bibinfo {author}
  {\bibfnamefont {M.}~\bibnamefont {Cabero}}, \bibinfo {author} {\bibfnamefont
  {D.~A.}\ \bibnamefont {Brown}}, \bibinfo {author} {\bibfnamefont {A.~H.}\
  \bibnamefont {Nitz}}, \ and\ \bibinfo {author} {\bibfnamefont
  {V.}~\bibnamefont {Raymond}},\ }\href {\doibase 10.1088/1538-3873/aaef0b}
  {\bibfield  {journal} {\bibinfo  {journal} {Publ. Astron. Soc. Pac.}\
  }\textbf {\bibinfo {volume} {131}},\ \bibinfo {pages} {024503} (\bibinfo
  {year} {2019})},\ \Eprint {http://arxiv.org/abs/1807.10312} {arXiv:1807.10312
  [astro-ph.IM]} \BibitemShut {NoStop}%
\bibitem [{\citenamefont {Ashton}\ \emph {et~al.}(2019)\citenamefont {Ashton},
  \citenamefont {Hübner}, \citenamefont {Lasky}, \citenamefont {Talbot},
  \citenamefont {Ackley}, \citenamefont {Biscoveanu}, \citenamefont {Chu},
  \citenamefont {Divakarla}, \citenamefont {Easter}, \citenamefont {Goncharov},
  \citenamefont {Vivanco}, \citenamefont {Harms}, \citenamefont {Lower},
  \citenamefont {Meadors}, \citenamefont {Melchor}, \citenamefont {Payne},
  \citenamefont {Pitkin}, \citenamefont {Powell}, \citenamefont {Sarin},
  \citenamefont {Smith},\ and\ \citenamefont {Thrane}}]{Ashton_2019}%
  \BibitemOpen
  \bibfield  {author} {\bibinfo {author} {\bibfnamefont {G.}~\bibnamefont
  {Ashton}}, \bibinfo {author} {\bibfnamefont {M.}~\bibnamefont {Hübner}},
  \bibinfo {author} {\bibfnamefont {P.~D.}\ \bibnamefont {Lasky}}, \bibinfo
  {author} {\bibfnamefont {C.}~\bibnamefont {Talbot}}, \bibinfo {author}
  {\bibfnamefont {K.}~\bibnamefont {Ackley}}, \bibinfo {author} {\bibfnamefont
  {S.}~\bibnamefont {Biscoveanu}}, \bibinfo {author} {\bibfnamefont
  {Q.}~\bibnamefont {Chu}}, \bibinfo {author} {\bibfnamefont {A.}~\bibnamefont
  {Divakarla}}, \bibinfo {author} {\bibfnamefont {P.~J.}\ \bibnamefont
  {Easter}}, \bibinfo {author} {\bibfnamefont {B.}~\bibnamefont {Goncharov}},
  \bibinfo {author} {\bibfnamefont {F.~H.}\ \bibnamefont {Vivanco}}, \bibinfo
  {author} {\bibfnamefont {J.}~\bibnamefont {Harms}}, \bibinfo {author}
  {\bibfnamefont {M.~E.}\ \bibnamefont {Lower}}, \bibinfo {author}
  {\bibfnamefont {G.~D.}\ \bibnamefont {Meadors}}, \bibinfo {author}
  {\bibfnamefont {D.}~\bibnamefont {Melchor}}, \bibinfo {author} {\bibfnamefont
  {E.}~\bibnamefont {Payne}}, \bibinfo {author} {\bibfnamefont {M.~D.}\
  \bibnamefont {Pitkin}}, \bibinfo {author} {\bibfnamefont {J.}~\bibnamefont
  {Powell}}, \bibinfo {author} {\bibfnamefont {N.}~\bibnamefont {Sarin}},
  \bibinfo {author} {\bibfnamefont {R.~J.~E.}\ \bibnamefont {Smith}}, \ and\
  \bibinfo {author} {\bibfnamefont {E.}~\bibnamefont {Thrane}},\ }\href
  {\doibase 10.3847/1538-4365/ab06fc} {\bibfield  {journal} {\bibinfo
  {journal} {The Astrophysical Journal Supplement Series}\ }\textbf {\bibinfo
  {volume} {241}},\ \bibinfo {pages} {27} (\bibinfo {year} {2019})}\BibitemShut
  {NoStop}%
\bibitem [{\citenamefont {Ahmed}\ \emph {et~al.}(2024)\citenamefont {Ahmed},
  \citenamefont {Kastha},\ and\ \citenamefont
  {Nielsen}}]{ahmed2024constraining}%
  \BibitemOpen
  \bibfield  {author} {\bibinfo {author} {\bibfnamefont {Z.}~\bibnamefont
  {Ahmed}}, \bibinfo {author} {\bibfnamefont {S.}~\bibnamefont {Kastha}}, \
  and\ \bibinfo {author} {\bibfnamefont {A.~B.}\ \bibnamefont {Nielsen}},\
  }\href@noop {} {\enquote {\bibinfo {title} {Constraining parametric deviation
  from kerr using black hole ringdown of gw150914 and gw190521},}\ } (\bibinfo
  {year} {2024}),\ \Eprint {http://arxiv.org/abs/2401.06049} {arXiv:2401.06049
  [gr-qc]} \BibitemShut {NoStop}%
\bibitem [{\citenamefont {Nishizawa}(2018)}]{Nishizawa_2018}%
  \BibitemOpen
  \bibfield  {author} {\bibinfo {author} {\bibfnamefont {A.}~\bibnamefont
  {Nishizawa}},\ }\href {\doibase 10.1103/physrevd.97.104037} {\bibfield
  {journal} {\bibinfo  {journal} {Physical Review D}\ }\textbf {\bibinfo
  {volume} {97}} (\bibinfo {year} {2018}),\
  10.1103/physrevd.97.104037}\BibitemShut {NoStop}%
\bibitem [{\citenamefont {D’Agostino}\ and\ \citenamefont
  {Nunes}(2019)}]{D_Agostino_2019}%
  \BibitemOpen
  \bibfield  {author} {\bibinfo {author} {\bibfnamefont {R.}~\bibnamefont
  {D’Agostino}}\ and\ \bibinfo {author} {\bibfnamefont {R.~C.}\ \bibnamefont
  {Nunes}},\ }\href {\doibase 10.1103/physrevd.100.044041} {\bibfield
  {journal} {\bibinfo  {journal} {Physical Review D}\ }\textbf {\bibinfo
  {volume} {100}} (\bibinfo {year} {2019}),\
  10.1103/physrevd.100.044041}\BibitemShut {NoStop}%
\bibitem [{\citenamefont {Finke}\ \emph {et~al.}(2021)\citenamefont {Finke},
  \citenamefont {Foffa}, \citenamefont {Iacovelli}, \citenamefont {Maggiore},\
  and\ \citenamefont {Mancarella}}]{Finke_2021}%
  \BibitemOpen
  \bibfield  {author} {\bibinfo {author} {\bibfnamefont {A.}~\bibnamefont
  {Finke}}, \bibinfo {author} {\bibfnamefont {S.}~\bibnamefont {Foffa}},
  \bibinfo {author} {\bibfnamefont {F.}~\bibnamefont {Iacovelli}}, \bibinfo
  {author} {\bibfnamefont {M.}~\bibnamefont {Maggiore}}, \ and\ \bibinfo
  {author} {\bibfnamefont {M.}~\bibnamefont {Mancarella}},\ }\href {\doibase
  10.1088/1475-7516/2021/08/026} {\bibfield  {journal} {\bibinfo  {journal}
  {Journal of Cosmology and Astroparticle Physics}\ }\textbf {\bibinfo {volume}
  {2021}},\ \bibinfo {pages} {026} (\bibinfo {year} {2021})}\BibitemShut
  {NoStop}%
\bibitem [{\citenamefont {Alfradique}\ \emph {et~al.}(2023)\citenamefont
  {Alfradique}, \citenamefont {Bom}, \citenamefont {Palmese}, \citenamefont
  {Teixeira}, \citenamefont {Santana-Silva}, \citenamefont {Drlica-Wagner},
  \citenamefont {Riley}, \citenamefont {Martínez-Vázquez}, \citenamefont
  {Sand}, \citenamefont {Stringfellow}, \citenamefont {Medina}, \citenamefont
  {Carballo-Bello}, \citenamefont {Choi}, \citenamefont {Esteves},
  \citenamefont {Limberg}, \citenamefont {Mutlu-Pakdil}, \citenamefont {Noël},
  \citenamefont {Pace}, \citenamefont {Sakowska},\ and\ \citenamefont
  {Wu}}]{alfradique2023dark}%
  \BibitemOpen
  \bibfield  {author} {\bibinfo {author} {\bibfnamefont {V.}~\bibnamefont
  {Alfradique}}, \bibinfo {author} {\bibfnamefont {C.~R.}\ \bibnamefont {Bom}},
  \bibinfo {author} {\bibfnamefont {A.}~\bibnamefont {Palmese}}, \bibinfo
  {author} {\bibfnamefont {G.}~\bibnamefont {Teixeira}}, \bibinfo {author}
  {\bibfnamefont {L.}~\bibnamefont {Santana-Silva}}, \bibinfo {author}
  {\bibfnamefont {A.}~\bibnamefont {Drlica-Wagner}}, \bibinfo {author}
  {\bibfnamefont {A.~H.}\ \bibnamefont {Riley}}, \bibinfo {author}
  {\bibfnamefont {C.~E.}\ \bibnamefont {Martínez-Vázquez}}, \bibinfo {author}
  {\bibfnamefont {D.~J.}\ \bibnamefont {Sand}}, \bibinfo {author}
  {\bibfnamefont {G.~S.}\ \bibnamefont {Stringfellow}}, \bibinfo {author}
  {\bibfnamefont {G.~E.}\ \bibnamefont {Medina}}, \bibinfo {author}
  {\bibfnamefont {J.~A.}\ \bibnamefont {Carballo-Bello}}, \bibinfo {author}
  {\bibfnamefont {Y.}~\bibnamefont {Choi}}, \bibinfo {author} {\bibfnamefont
  {J.}~\bibnamefont {Esteves}}, \bibinfo {author} {\bibfnamefont
  {G.}~\bibnamefont {Limberg}}, \bibinfo {author} {\bibfnamefont
  {B.}~\bibnamefont {Mutlu-Pakdil}}, \bibinfo {author} {\bibfnamefont
  {N.~E.~D.}\ \bibnamefont {Noël}}, \bibinfo {author} {\bibfnamefont {A.~B.}\
  \bibnamefont {Pace}}, \bibinfo {author} {\bibfnamefont {J.~D.}\ \bibnamefont
  {Sakowska}}, \ and\ \bibinfo {author} {\bibfnamefont {J.~F.}\ \bibnamefont
  {Wu}},\ }\href@noop {} {\enquote {\bibinfo {title} {A dark siren measurement
  of the hubble constant using gravitational wave events from the first three
  ligo/virgo observing runs and delve},}\ } (\bibinfo {year} {2023}),\ \Eprint
  {http://arxiv.org/abs/2310.13695} {arXiv:2310.13695 [astro-ph.CO]}
  \BibitemShut {NoStop}%
\end{thebibliography}%

\end{document}